Enhancement of the magnetocaloric effect driven by changes in the crystal structure of Al-doped GGG, $Gd_3Ga_{5-x}Al_xO_{12}$ ($0 \leq x \leq 5$)

A. C. Sackville Hamilton[1], G. I. Lampronti[2], S. E. Rowley[1,3], and S. E. Dutton[1*]

[1]Cavendish Laboratory, University of Cambridge, JJ Thomson Avenue, Cambridge, CB3 0HE, UK

[2]Department of Earth Sciences, University of Cambridge, Downing Street, Cambridge, CB2 3EQ, UK

[3]Centro Brasileiro de Pesquisas Físicas, Rua Dr Xavier Sigaud 150, Urca, Rio de Janeiro, 22290-180, Brasil

[*]corresponding author: sed33@cam.ac.uk

**Abstract**

The $Gd_3Ga_{5-x}Al_xO_{12}$ ($0 \leq x \leq 5$) solid solution has been prepared using ceramic synthesis routes and the structural and magnetic properties investigated using X-ray diffraction, magnetic susceptibility, $\chi$, and isothermal magnetisation, M(H), measurements. Our results indicate a contraction of the unit cell and more significant antiferromagnetic interactions as x increases. Despite the decrease in the magnetic polarisation on application of a field and the corresponding decrease in the change in the magnetic entropy we find that $Gd_3Al_5O_{12}$ has a significantly higher observed (17%) and theoretical (14%) $\Delta S$ per unit mass than $Gd_3Ga_5O_{12}$. Per unit volume the theoretical increase in $\Delta S$ (7%) is offset by the increased antiferromagnetic interactions in $Gd_3Al_5O_{12}$. The differences in $\Delta S$ are driven by a decrease in both the mass and density as Al ions replace Ga ions. These results highlight the importance of changes to the crystal structure when considering materials for solid state magnetic cooling.

# 1. Introduction

Magneto- (or electro-) caloric cooling are both viable alternatives to cooling using liquid cryogens.(1) In materials of this kind application of a magnetic (or electric) field induces a change in entropy of a system which when followed by adiabatic demagnetisation induces a reduction in temperature, $\Delta T$.(2) In most materials the change in entropy, $\Delta S$, and temperature, $\Delta T$, is maximised close to a phase transition, $T_{order}$, and hence for practical applications a material will tend to operate close to this. At high temperatures a wide range of materials have been identified as potential magnetocalorics,(3) however identifying suitable materials for cryogen refrigeration is more challenging. At low temperatures a restriction on a materials cooling ability is that the operating temperature range should be greater than the magnetic ordering $T$, $T_{order}$. Materials with low $T_{order}$ used in magnetic refrigeration include cerium magnesium nitrate(4, 5) (CMN), $Ce(NO_3)_3 \cdot 3Mg(NO_3)_2 \cdot 24H_2O$ and ferric ammonium alum(5) (FAA), $FeNH_4(SO_4)_2 \cdot 12H_2O$. These have dilute magnetic lattices to suppress the magnetic ordering and allow for cooling to sufficiently low $T$. Whilst effective, poor thermal conductivity necessitates the use of complex assemblies to ensure sufficient thermal contact in working devices. Sensitivity to elevated $T$ and chemical reactivity also complicate handling of such materials.(5)

Rather than using a dilute magnetic system, an alternative way to suppress the ordering transition of a magnetic material is through frustration of the magnetic lattice. Geometrically frustrated magnets (GFMs) have a magnetic lattice in which, due to its geometry, it is impossible to completely satisfy all of the magnetic interactions simultaneously. An ordered magnetic ground state thus only occurs as a result of a delicate balance between competing magnetic interactions. In GFMs the magnetic ordering transition, $T_{order}$, is thus suppressed to significantly below the Weiss Temperature, $\theta$. For lanthanide ions, $Ln^{3+}$, where $\theta$ is already small as a result of the highly localised $4f$ orbitals the magnetic ordering transition can be suppressed to temperatures approaching, or even below, those of CMN and FAA.(6) Of particular interest for its large MCE is gadolinium gallium garnet (GGG), $Gd_3Ga_5O_{12}$.(1, 7) GGG is a GFM containing isotropic $Gd^{3+}$ which has a large effective magnetic moment ($\mu_{eff}$ = 7.98 $\mu_B$ per $Gd^{3+}$). The frustration of the magnetic lattice in GGG results in a complex magnetic phase diagram below the antiferromagnetic ordering $T$, $T_N$ which is still the subject of debate.(8-12) Recent results have indicated that the low $T$ magnetic properties of Gd-containing garnets are highly dependent on the occupancy of the non-magnetic sites.(12) In garnets the cation sites can be divided into three distinct environments, an eight co-ordinate site, occupied by $Gd^{3+}$ in GGG, and octahedral and tetrahedral sites occupied by $Ga^{3+}$ in GGG. Comparison of the low $T$ ground states of $Gd_3Ga_5O_{12}$, $Gd_3Al_5O_{12}$ and $Gd_3Te_2Li_3O_{12}$ indicates a subtle interplay between the nearest neighbour magnetic exchange interaction, $J_1$, and the dipolar interaction, $D$.(12) The impact these changes may have on the higher $T$ properties and the MCE have yet to be explored. Overall the impact of doping on the MCE in GGG, has been relatively under-investigated. However doping of $Gd^{3+}$ with $Tb^{3+}$ or $Dy^{3+}$ and inclusion of $Fe^{3+}$ on the $Ga^{3+}$ sites have both been demonstrated to have a significant impact on the measured MCE. [13-16]

In this paper we describe the effect of replacing the gallium on both the tetrahedral and octahedral sites either partially or completely with aluminium. Our results indicate that whilst the volume contraction on replacing $Ga^{3+}$ with smaller $Al^{3+}$ ions is linear, the relative compressibility of the octahedral and tetrahedral sites results in preferential occupation of the tetrahedral sites with gallium and a change in the crystal electric field of the gadolinium ions. We find that above 2 K the magnetic susceptibility and isothermal magnetisation is suppressed as the aluminium concentration increases. The effect on the MCE on doping is subtle; per $mol_{Gd}$ the MCE decreases as aluminium replaces gallium, however the reduced molar mass and density results in an increase in the MCE per kg as more $Al^{3+}$ is introduced into GGG. The change in entropy per unit volume remains constant. Our results indicate that Al-doped GGG and $Gd_3Al_5O_{12}$ (GAG) may thus be more effective for magnetic cooling, at least for $T > 2$ K, than undoped GGG.

## 2. Experimental

Samples of $Gd_3Ga_{5-x}Al_xO_{12}$ ($0 \leq x \leq 5$) were prepared using a ceramic synthesis method. For $0 \leq x \leq 3$ samples were prepared by mixing stoichiometric amounts of pre-dried Gadolinium oxide (99.999%, Alfa Aesar), pre-dried Gallium oxide (99.999%, Alfa Aesar) and Aluminium oxide (99.995%, Alfa Aesar). Pressed pellets were then heated to 1275 °C or 1200 °C in air for 48 hours for the x = 0 and x = 2, 3 samples respectively. This heating procedure was continued until powder X-ray diffraction indicated the formation of a phase pure product. Attempts to prepare $Gd_3Ga_{5-x}Al_xO_{12}$ with x > 3 using this method resulted in the formation of large amounts of a $GdAlO_3$ impurity phase. To prepare $Gd_3Al_5O_{12}$ an alternative synthesis route was used in which stoichiometric amounts of pre-dried Gadolinium nitrate hydrate (99.99%, Alfa Aesar), pre-dried Aluminum nitrate nonahydrate (99.999%, Alfa Aesar) and pre-dried Gallium oxide were intimately mixed. To allow for decomposition of the nitrate starting materials, a low temperature heating step at 1000 °C for 12 hours in air was carried out prior to heating at successively high temperatures, $T_{max}$ = 1275 °C, for 48 hours in air. This final heating step was repeated until powder X-ray diffraction indicated formation of a phase pure product. For all heat treatments, samples were heated and cooled at 3 °C / min.

The progress of the reaction was monitored with short scans, $5 \leq 2\theta \leq 60°$, using a Panalytical Empyrean X-ray diffractometer with Cu K$\alpha$ radiation. High resolution data for quantitative analysis of the crystal structure were collected using a Bruker D8 Advance diffractometer operating with Cu K$\alpha$ radiation, a Ge monochromator and a Sol-XE energy dispersive detector. Measurements were made over a wide angular range, $10 \leq 2\theta \leq 120°$, with $\Delta 2\theta = 0.02°$ over ~18 hours. Rietveld analysis(17) was carried out using the *GSAS*(18) suite of programs with the *EXPGUI* interface.(19) Backgrounds were fitted using a Chebyshev polynomial of the first kind and the peak shape modelled using a pseudo-Voigt function.

Magnetic measurements, $2 \leq T \leq 300$ K, were made on a Quantum Design Magnetic Properties Measurement System (MPMS) with a Superconducting Quantum Interference Device (SQUID) after cooling in both zero field (ZFC) and the 100 Oe measuring field (FC). At selected temperatures isothermal magnetisation measurements, $0 \leq \mu_o H \leq 5$ T, were made on application and removal of the applied field after cooling in zero field.

3. Results and Discussion

3.1. Structural Properties of $Gd_3Ga_{5-x}Al_xO_{12}$ ($0 \leq x \leq 5$)

The $Gd_3Ga_{5-x}Al_xO_{12}$ ($0 \leq x \leq 5$) solid solution was successfully prepared using a solid state synthesis route. The details of the Rietveld analysis are given in Table 1 and Fig. 1b. The crystal chemistry of lanthanide garnets is complex and non-stoichiometry due to excess lanthanide ions occupying the octahedral and tetrahedral sites has been documented to have a dramatic effect on the observed properties.(11) In GGG it is possible to quantify this nonstoichiometry(20, 21) using the parameter y in $Gd_{3+y}Ga_{5-y}O_{12}$, by looking at how far the lattice parameter deviates from the ideal value of 12.375 Å. In the sample of GGG measured here we calculate that y = 0.0169(9), corresponding to a 0.56(3) % excess of $Gd^{3+}$. For the samples containing aluminium no such model exists, however the presence of small amounts of $GdAlO_3$ suggests that there may be a small Gd excess in these samples. Any excess present is small and beyond the resolution of our X-ray measurements, placing an upper limit of 0.4 % Gd excess. The presence of the GdAlO$_3$ impurity phase in the mixed Ga/Al ion systems also raises the possibility of inhomogeneity in the ratio of Ga:Al, we estimate this to be < 5% excess of Ga.

As expected from Vegard's law a monotonic decrease in the lattice parameter is observed as Ga$^{3+}$ is replaced with smaller Al$^{3+}$ ions in Gd$_3$Ga$_{5-x}$Al$_x$O$_{12}$ (0 ≤ x ≤ 5) (Fig. 2a). Despite the linear decrease in lattice parameter, closer consideration of the results of the Rietveld analysis indicate that substitution of Ga$^{3+}$ for Al$^{3+}$ results in subtle changes to the crystal structure (Fig. 2, Tables 1-2). The size of the octahedral and tetrahedral sites decrease monotonically with x (Fig. 2d and Table 2). However, the decrease in size is significantly larger for the tetrahedral (5.9%) compared to the octahedral sites (1.6%). This change in the relative size of the two sites is driven by the larger difference in size between the ionic radii of 4-coordinate Ga$^{3+}$ and Al$^{3+}$(22) and facilitated by displacement of the O$^{2-}$ sublattice. The change in the oxygen lattice can be seen in the shift in the Gd-O environment as a function of composition (Fig. 2c and Table 2). In the two end members, GGG and GAG, there is no significant difference in the Gd-O bond lengths, but in the intermediate compositions a significant increase in the shorter of the Gd-O interactions is observed, rendering the Gd$^{3+}$ crystal electric field (CEF) more isotropic. We ascribe this change in the gadolinium environment to be a consequence of the changing size of the Ga/Al sites. Unexpectedly, given the linearity in the decrease of both the lattice parameter and Ga/Al-O bond lengths, a significant preference for Ga to occupy the tetrahedral sites is observed in the mixed Ga-Al compositions. The reason for this is unclear, but may reflect a preference for ordering of the Ga and Al ions onto the tetrahedral and octahedral sites respectively. Experiments in which the samples are cooled more slowly to allow for formation of the thermodynamic rather than the kinetic product are required to test this hypothesis.

3.2. Magnetic Properties Gd$_3$Ga$_{5-x}$Al$_x$O$_{12}$ (0 ≤ x ≤ 5)

The magnetic susceptibility, $\chi$ = M/H, and its inverse, $\chi^{-1}$, for Gd$_3$Ga$_{5-x}$Al$_x$O$_{12}$ (0 ≤ x ≤ 5) are presented in Fig. 3a. Within the *T* range measured no sign of magnetic ordering is observed in any of the compositions and no irreversibility between the ZFC and FC measurements is detected. As x increases, a decrease in the magnetic susceptibility at low *T* is registered. The inverse susceptibility is linear across the entire *T* range and fitting to the Curie-Weiss Law, $\chi$ = C/(*T*-$\theta$) (C = $\frac{N\mu_{eff}^2}{3k_B}$ = Curie constant, $\theta$ = Weiss temperature) was carried out for $T > 2$ K. The parameters from the fit are included in Table 3. For all compositions the effective magnetic moment, $\mu_{eff}$, is close to the theoretical value for $Gd^{3+}$ of 7.98 $\mu_B$. As the aluminium content is increased the magnitude of $\theta$ increases from -2.6 K in GGG to -4.0 K in GAG. At intermediate compositions $\theta$ is constant at ~-3.1 K. This indicates that the magnitude of the antiferromagnetic (AFM) interactions increase as Al is doped into GGG.

Isothermal magnetisation measurements for $Gd_3Ga_{5-x}Al_xO_{12}$ ($0 \leq x \leq 5$) are shown in Fig. 3b. For all compositions no hysteresis between the application and removal of the field is observed. At $T \leq 10$ K the magnetic moment begins to saturate in the limiting field of 5 T. As $T$ decreases the impact of field on the magnetisation becomes more significant. The impact of changes in composition is small but for $T \leq 10$ K a clear difference between the different compositions is observed. As the concentration of aluminium in GGG increases the polarisability in a field decreases, mirroring the trend in $\theta$ and the stronger antiferromagnetic interactions in the Al-containing samples.

Previous studies on GGG have considered a number of factors which affect the magnetic behaviour, including the strength of the nearest neighbour interaction, $J_1$, the dipolar interaction, $D$, and the single ion anisotropy, $\Delta$.(12) In $Gd_3Ga_{5-x}Al_xO_{12}$ ($0 \leq x \leq 5$) the absolute and relative magnitude of these parameters would be expected to change as a function of composition. From our analysis of the structure and magnetic susceptibility it is possible to approximate changes in $J_1$ and $D$ (Table 3). As x increases, we find that the absolute value of both $J_1$ and $D$ increases, but that the nearest neighbour exchange increasingly dominates the dipolar interaction. Changes in the single ion anisotropy, $\Delta$, are harder to quantify without more detailed measurements. $\Delta$ for GGG is known to be small ($\Delta < 40$ mK(23)), consistent with the isotropic $5f^7$ electronic configuration. On doping, the small changes in the Gd-O environment may result in a slight decrease in $\Delta$ as the site becomes more symmetric, but overall these changes are anticipated to be minimal. To fully understand how the changes in $J_1$, $D$ and $D/J_1$ impact the magnetic properties, measurements to lower $T$ are required. However, in our measurements the increase in both $J_1$ and $D$ is revealed by the suppression of both $\chi_{2K}$ and the isothermal magnetisation, $4 \leq T \leq 10$ K, when Ga is substituted for Al. The wide $T$ and $\mu_0 H$ range over which these AFM fluctuations are present may support the hypothesis, presented elsewhere, that the interplay between $J_1$ and $D$ has a large effect on the magnetic ground state.(12)

### 3.3. Magnetocaloric Effect of $Gd_3Ga_{5-x}Al_xO_{12}$ ($0 \leq x \leq 5$)

The magnetocaloric effect (MCE) for $Gd_3Ga_{5-x}Al_xO_{12}$ ($0 \leq x \leq 5$) was estimated from the isothermal magnetisation measurements assuming that the entropy change, $\Delta S$, on application of a field is given by the Maxwell relationship, $\Delta S = \int_0^H \left(\frac{dM}{dT}\right)_H dH$. The results, depicted graphically in Fig. 4, show a large variability dependant on the units in which $\Delta S$ is considered. If $\Delta S$ is presented per mol of Gd ions (Fig. 4a) then the trends in $\chi_{2K}$ and the isothermal magnetisation are reflected in the MCE. Intuitively this reflects the larger polarisation, and hence entropy change, possible in GGG when compared with the Al-containing samples on application of a magnetic field. For all compositions the strength of the applied field has a dramatic effect on $\Delta S$, in the limits of our experiments, $T = 4$ K and $\mu_0H = 5$ T, $\Delta S$ is approaching half of the value anticipated from the magnetic moments, $R\ln(2J+1) = 17.28$ J K$^{-1}$ / mol$_{Gd}^{-1}$. This indicates that at lower $T$ there may still be significant entropy for magnetic cooling.

Whilst presenting $\Delta S$ per mol$_{Gd}$ is informative for understanding how close to the theoretical limits our experiments are, for practical applications $\Delta S$ per unit mass or volume is more useful as these give an indication of the quantity of material required. As Al is added to GGG both the volume per formula unit and density decrease, changing the trend observed for $\Delta S$ per mol$_{Gd}$. For the theoretical change in magnetic entropy (Table 4) the relative changes in $\Delta S$, defined by ($\left[\frac{\Delta S_{GAG}-\Delta S_{GGG}}{\Delta S_{GGG}}\right]$ x100), for GGG and GAG are 0, 14 and 7 % in units of J K$^{-1}$ mol$_{Gd}^{-1}$, J K$^{-1}$ kg$^{-1}$ and J K$^{-1}$ cm$^{-3}$ respectively. This is also reflected in the divergence of the calculated $\Delta S$ per unit mass for our measurements where a 17 % increase is observed in GAG relative to GGG (Fig. 4b). However, when $\Delta S$ per unit volume is considered, the entropy gain due to the reduced unit cell is offset by the magnetic ordering such that $\Delta S$ per cm$^3$ for all samples is essentially constant (Fig. 4c). The theoretical $\Delta S$ per unit volume surpass those of CMN and FAA (Table 4) by an order of magnitude, reflecting the advantages of using GFM systems for magnetic cooling. Furthermore the advantages of reducing the mass and the density of the unit cell are reflected in the enhancement of the maximum entropy change per unit volume and mass in Al-doped GGG. With suitable engineering to allow for sufficient thermal transport Al-doped GGG or GAG may be viable alternatives to FAA and CMN for solid state magnetic cooling using the MCE. Further experiments at low $T$ are required to confirm this and to find the limiting cooling $T$, as defined by the magnetic ordering $T$.

## 4. Conclusion

Our results indicate the importance of changes to the structure when evaluating materials for magnetic refrigeration. In the $Gd_3Ga_{5-x}Al_xO_{12}$ ($0 \leq x \leq 5$) solid solution discussed here, the impact is three-fold, with the changes to the volume, density of magnetic ions and magnetic ordering all impacting the observed change in entropy. If the theoretical magnetic entropy change of GAG and its derivatives can be realised the impact of changing the structure is dramatic. When searching for GFM materials for magnetic refrigeration we anticipate that the interplay between the crystal and magnetic structure will be crucial to understanding the optimal composition of a given system.

## 5. Acknowledgements

We acknowledge assistance from S. Redfern during our X-ray diffraction measurements. The authors thank C. S. Haines, C. Lui, S. Saxena and G. Lonzarich for their helpful discussions. SER acknowledges funding from Emmanuel College, Cambridge. This work was supported by the Winton Programme for the Physics of Sustainability

**Figure Captions**

Figure 1: a) Garnet crystal structure (left). Gadolinium sites are shown in white, the tetrahedral and octahedral Gallium/Aluminium positions are shown in orange and blue respectively. b) X-ray diffraction pattern for $Ga_3Gd_3Al_2O_{12}$. Collected data (+), modelled data (black line) and difference (blue line) are shown. The tick marks are for the main phase (upper) and $GdAlO_3$ (lower) impurity phase (4.6(2) wt%).

Figure 2: Variation of a) lattice parameter, b) fractional occupancy of Ga/Al sites, c) Gd-O bond lengths and d) Ga/Al-O bond lengths as a function of x for $Gd_3Ga_{5-x}Al_xO_{12}$. In c) a structural model of the Gd-O coordination environment is inset, blue and green bonds correspond to the short and long Gd-O bonds respectively.

Figure 3: a) Magnetic susceptibility and inverse (inset) as a function of temperature and b) Isothermal magnetisation as a function of field at selected temperatures for $Gd_3Ga_{5-x}Al_xO_{12}$.

Figure 4: Change in magnetic entropy, $\Delta S$, a) per $mol_{Gd}$ b) per kg and c) per $cm^3$ as a function of temperature in 1 and 5 T fields for $Gd_3Ga_{5-x}Al_xO_{12}$. The change in entropy as a function of applied field at 4 K is inset.

Table 1: Structural parameters for $Gd_3Ga_{5-x}Al_xO_{12}$ as obtained from Rietveld analysis of X-ray diffraction measurements at room temperature. Refinements were carried out in the $Ia\bar{3}d$ space group, with Gd on the 24c site (1/8,0,1/4), Ga/Al1 on the 16a (0,0,0), Ga/Al2 24d (3/8,0,1/4) sites and O on a 96h (x,y,z) position.

| $Gd_3Ga_{5-x}Al_xO_{12}$ | $Ia\bar{3}d$ | x = 0 | x = 2 | x = 3 | x = 5 |
|---|---|---|---|---|---|
| a (Å) | | 12.3789(2) | 12.2707(3) | 12.2233(2) | 12.1144(2) |
| V (Å$^3$) | | 1896.91(9) | 1847.62(11) | 1826.29(7) | 1777.88(7) |
| $\chi^2$ | | 0.7065 | 1.263 | 1.223 | 1.262 |
| Ga/Al1 | Frac Ga | 1.0 | 0.436(12) | 0.232(11) | 0.0 |
| Ga/Al2 | Frac Ga | 1.0 | 0.709(8) | 0.512(7) | 0.0 |
| O | x<br>y<br>z | 0.9721(12)<br>0.0491(9)<br>0.1482(15) | 0.9648(9)<br>0.0489(10)<br>0.14634(10) | 0.9650(8)<br>0.0460(9)<br>0.1469(9) | 0.9657(8)<br>0.0461(9)<br>0.1489(9) |
| $B_{ov}$ (Å$^2$) | | 0.0 | 0.0 | 0.0 | 0.0055(6) |
| wt% GdAlO$_3$ | | - | 4.6(2) | 2.9(2) | 3.6(2) |
| Wt% Al$_2$O$_3$ | | - | - | - | 3.5(2) |

Table 2: Bonds lengths for $Gd_3Ga_{5-x}Al_xO_{12}$ as obtained from Rietveld analysis of X-ray diffraction measurements at room temperature.

| $Gd_3Ga_{5-x}Al_xO_{12}$ | x = 0 | x = 2 | x = 3 | x = 5 |
|---|---|---|---|---|
| Gd-Gd (Å) | 3.79023(4) x4 | 3.75714(6) x4 | 3.74262(3) x 4 | 3.70925(4) x4 |
| Gd-O (Å)<br><br><Gd-O> (Å) | 2.353(18) x4<br>2.526(19) x4<br>2.440 | 2.417(12) x4<br>2.519(13) x4<br>2.468 | 2.394(10) x4<br>2.543(11) x4<br>2.469 | 2.353(11) x4<br>2.521(12) x4<br>2.437 |
| Ga/Al1-O (Å) | 1.965(18) x 6 | 1.943(13) x6 | 1.930(11) x6 | 1.934(11) x 6 |
| Ga/Al2-O (Å) | 1.844(16) x4 | 1.786(12) x4 | 1.765(10) x4 | 1.738(10) x 4 |
| <Ga/Al-O> (Å) | 1.92 | 1.88 | 1.86 | 1.86 |

Table 3: Magnetic parameters for $Gd_3Ga_{5-x}Al_xO_{12}$ obtained from fitting to the Curie Weiss law, $T < 30$ K. The nearest neighbour interaction, $J_1$, was defined using the relationship, $\theta = \frac{J_1 nS(S+1)}{3k_B}$ where the coordination number n = 4 and the dipolar interaction, $D = \frac{\mu_0 \mu_{eff}^2}{4\pi R_{nn}^3}$ where $R_{nn}$ is the nearest neighbour Gd-Gd interaction.

| $Gd_3Ga_{5-x}Al_xO_{12}$ | θ (K) | $\mu_{eff}$ ($\mu_B$) | $J_1$ (mK) | D (mK) | $D/J_1$ |
|---|---|---|---|---|---|
| 0 | -2.6(1) | 8.0(1) | 124(4) | 45(1) | 0.36 |
| 2 | -3.1(1) | 7.8(1) | 148(4) | 46(1) | 0.31 |
| 3 | -3.2(1) | 7.9(1) | 152(4) | 47(1) | 0.31 |
| 5 | -4.0(1) | 7.7(1) | 190(4) | 48(1) | 0.25 |

Table 4: Theoretical maxima for change in magnetic entropy, $-\Delta S_{max}$ = Rln(2J+1), for $Gd_3Ga_{5-x}Al_xO_{12}$ in a variety of units. *given in units of J $K^{-1}$ $mol^{-1}$ (both FAA and CMN contain a single magnetic ion per formula unit).

| $Gd_3Ga_{5-x}Al_xO_{12}$ | $-\Delta S_{max}$ | | |
|---|---|---|---|
| x | (J $K^{-1}$ $mol_{Gd}^{-1}$) | (J $K^{-1}$ $kg^{-1}$) | (J $K^{-1}$ $cm^{-3}$) |
| 0 | Rln(2J+1) = 17.28 | 51.2 | 0.363 |
| 2 | 17.28 | 55.9 | 0.373 |
| 3 | 17.28 | 58.6 | 0.377 |
| 5 | 17.28 | 64.9 | 0.387 |
| FAA | 14.89* | 30.9 | 0.034 |
| CMN | 14.89* | 14.6 | 0.031 |

Figure 1

a)

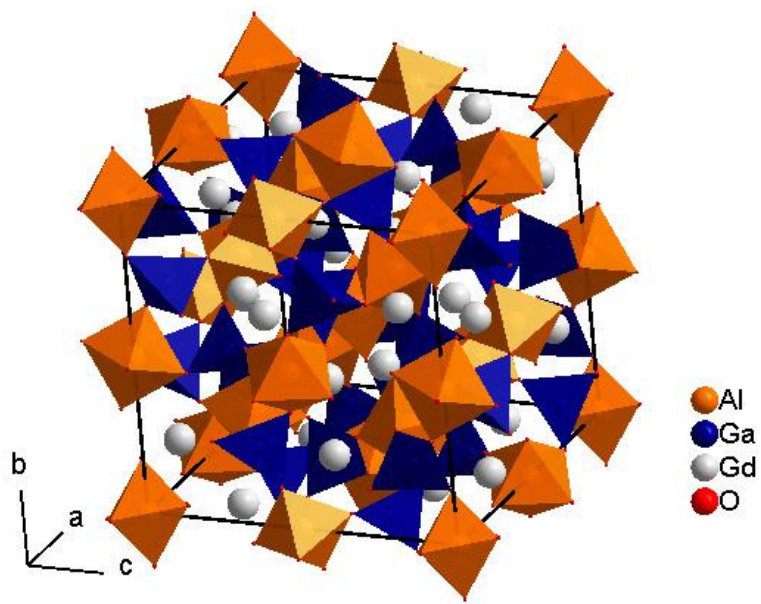

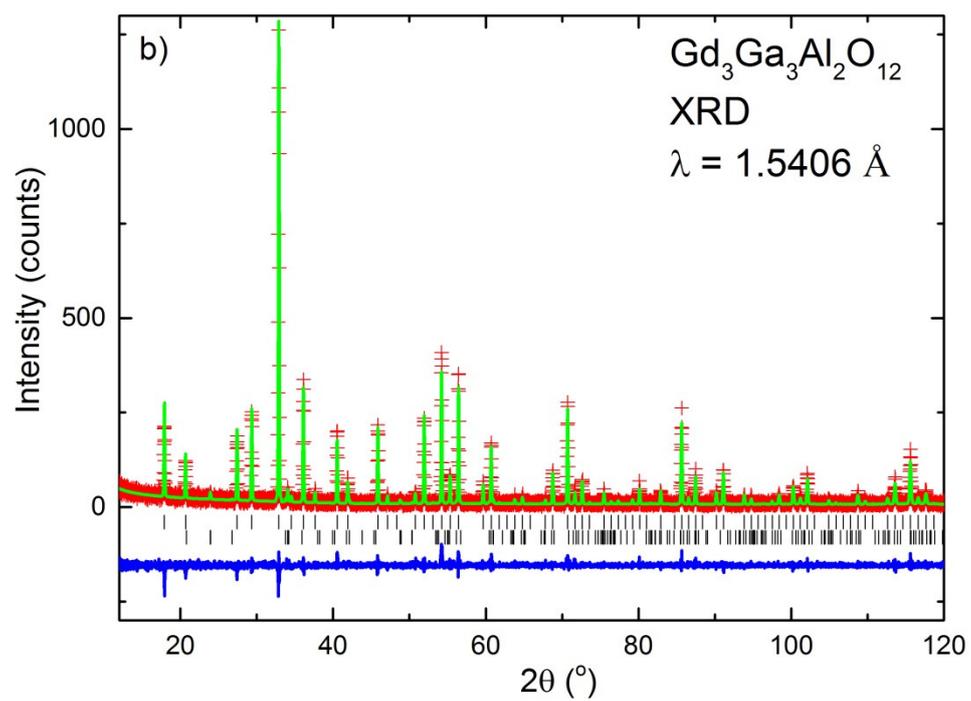

Figure 2

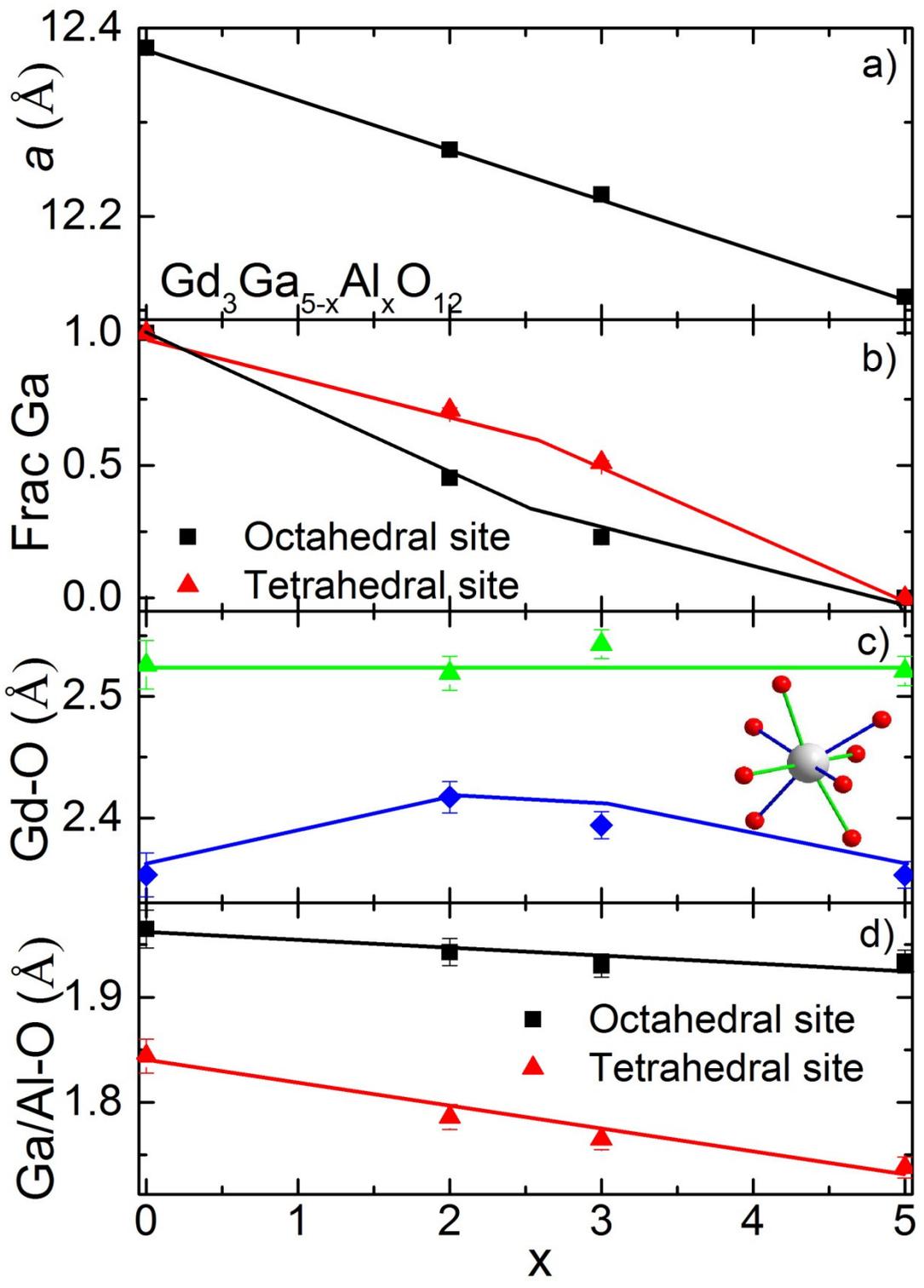

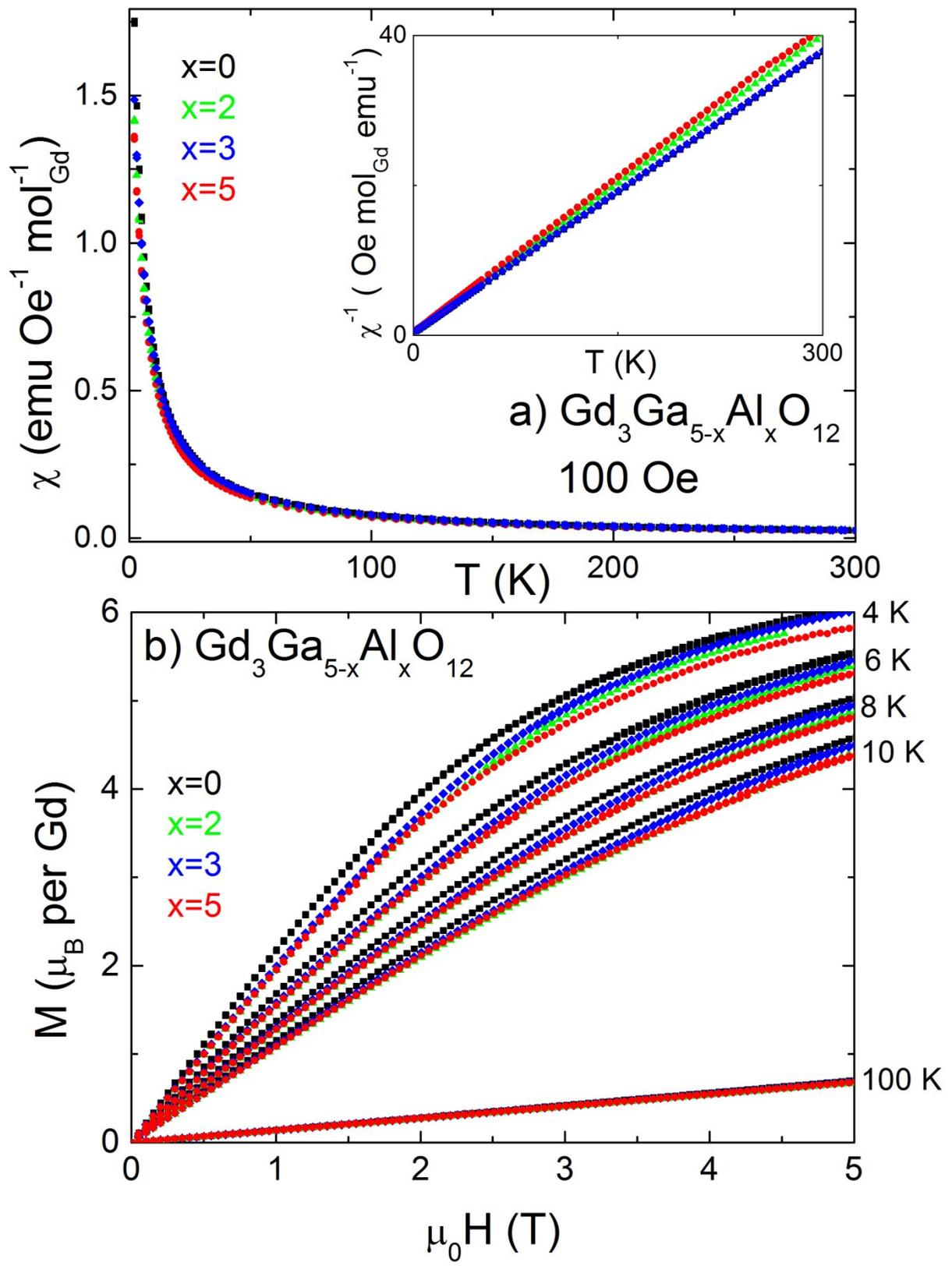

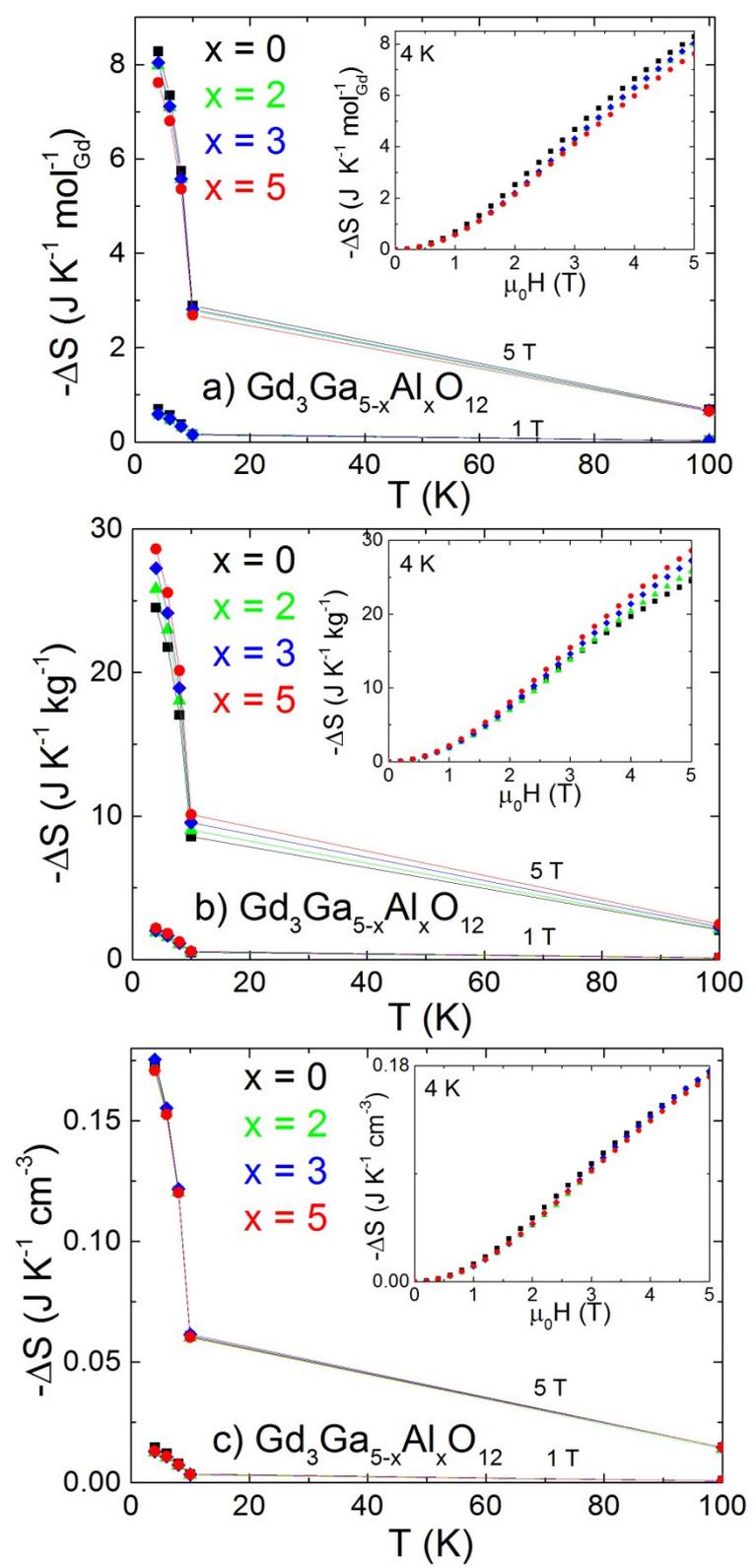